\documentclass[aps,tightenlines,preprint,superscriptaddress,showpacs,%
floatfix,amssymb,byrevtex,nofootinbib]{revtex4}  

\usepackage{epsfig,bm,dcolumn}

\begin{document}

\preprint{hep-ph/0605105}


\title{Scalar $\bm{\kappa}$ meson in $\bm{K^*}$ photoproduction}


\author{Yongseok Oh}%
\email{yoh@physast.uga.edu}

\affiliation{Department of Physics and Astronomy,
University of Georgia, Athens, Georgia 30602, U.S.A.}

\author{Hungchong Kim}%
\email{hungchon@postech.ac.kr}

\affiliation{Department of Physics, Pohang University of Science and
Technology, Pohang 790-784, Korea}



\begin{abstract}

We propose that the scalar $\kappa(800)$ meson may play an important role
in $K^*$ photoproduction.
In the reactions of $\gamma p \to K^{*+} \Lambda$ and
$\gamma p \to K^{*0} \Sigma^+$, we consider the production mechanisms
including $t$-channel $K^*$, $K$, $\kappa$ exchanges, $s$-channel $N$,
$\Delta$ diagrams, and $u$-channel $\Lambda$, $\Sigma$, $\Sigma^*$ diagrams
within the tree level approximation, and find that the $\kappa$-meson
exchange may contribute significantly to $K^*\Sigma$ photoproduction,
while it is rather supplementary in $K^*\Lambda$ photoproduction.
We demonstrate how the observables of $K^*$ photoproduction can be used to
constrain the $\kappa$ meson properties.
In particular, the parity asymmetry can separate the $\kappa$
meson contribution in $K^*$ photoproduction.

\end{abstract}


\maketitle

Recently, the CLAS Collaboration at Thomas Jefferson National
Accelerator Facility reported preliminary cross section data for $K^*(892)$
photoproductions, namely, $\gamma p \to K^{*0} \Sigma^+$~\cite{HH05b}
and $\gamma p \to K^{*+} \Lambda$~\cite{GW06}.
In the baryon sector, $K^*$ vector meson photoproduction 
can be used to search for the nucleon resonances which couple strongly
to the $K^* Y$ channel, where $Y$ stands for a hyperon~\cite{CR98b}.
This reaction is interesting in the meson sector as well since it can
offer an opportunity to study the scalar $\kappa(800)$ meson whose 
exchange is prohibited in $K$ meson photoproduction.

Since the Pomeron exchange is absent in the photoproduction of strange
mesons, the main production mechanisms of $K^*$ photoproduction
should be different from the case of non-strange neutral vector mesons
($\rho^0,\omega,\phi$)~\cite{OTL01-OL04}.
In Ref.~\cite{ZAB01}, Zhao {\it et al.\/} have studied $K^*\Sigma$
photoproduction within a quark model.
Some assumptions were made on the quark-meson couplings and parameters,
which should be further tested by experiments.
We have studied $\gamma N \to K^* \Lambda$ reaction in
Ref.~\cite{OK06a}, and found that the $t$-channel $K$ exchange
dominates the production amplitudes at small scattering angles
and it can describe quite well the total cross section
data of Ref.~\cite{GW06}.

The two preliminary experimental data of CLAS for $K^{*+}\Lambda$ and
for $K^{*0}\Sigma^+$ photoproductions~\cite{HH05b,GW06} show a very
challenging aspect that requires careful examinations.
Namely, the two production processes have very similar cross sections,
not only in the magnitude but also in the angular distribution at forward
scattering region~\cite{HG}.
This contradicts with a naive expectation based on the kaon
exchange process which predicts that the cross section for
$K^{*+}\Lambda$ production would be larger than that for
$K^{*0}\Sigma^+$ production by a factor of $\sim 3$, since
$R_K \equiv (g^c_{K^*K\gamma} g_{KN\Lambda}^{} / \sqrt2 g^0_{K^*K\gamma}
g_{KN\Sigma}^{})^2  = [g^c_{K^*K\gamma}(1+2\alpha)/\sqrt{6}
g^0_{K^*K\gamma} (1-2\alpha)]^2 \simeq 1.7^2$
with $\alpha = f/(f+d) \approx 0.365$~\cite{RSY99-SR99}.
(Here $\sqrt2$ is the isospin factor.)
To compensate this difference, it is necessary to have different
production mechanisms for $K^*\Sigma$ production from the $K^*\Lambda$
production case, unless we assume a large value of $g_{KN\Sigma}^{}$
to have $R_K \sim 1$.
Sizable $s$-channel nucleon resonance effects, which could
be responsible for the similarities between $K^+\Lambda$ and
$K^+\Sigma^0$ photoproductions at low energies~\cite{CLAS05c},
are not sufficient to explain the similarities in 
$K^*$ photoproductions at {\it forward\/} angles with relatively high
energies.
In order to have similar differential cross sections at forward angles,
we expect to have other $t$-channel mechanisms that contribute
significantly to $K^*\Sigma$ production but give supplementary contribution
to $K^*\Lambda$ production. 
In this paper, we propose that the light scalar $\kappa(800)$ meson 
can have this role, which can actually explain the observed
similarities between the cross sections for
$\gamma p \to K^{*+} \Lambda$ and for $\gamma p \to K^{*0} \Sigma^+$.

The nature of the scalar mesons is yet to be clarified and there are many
models on the structure of scalar meson nonet~\cite{Jaf77a-WI90-CDGKR93}.
In the case of scalar $\kappa(800)$ meson, the situation is even
worse since its existence is still controversial~\cite{PDG04} as can be
seen in many pros and cons
\cite{Oller03b-BFSS98-Bugg03-CP01,BES05a-E791-02-FOCUS-02-CLEO-01b,%
Ishida04-Bugg04d-AT04}.
Accordingly, the predicted or estimated mass and width of the $\kappa$ are in
a broad range:
$M_\kappa = 600 \sim 900$ MeV and $\Gamma_\kappa = 400 \sim 770$ MeV
\cite{PDG04}.
Here, we do not address the issue whether such a light $\kappa$ exists
in nature,
but instead we demonstrate how one can explain the similarities observed
in $K^*$ photoproductions by introducing light $\kappa$ meson and how one
can identify its role through some observables of this reaction.

\begin{figure}[t]
\centering
\epsfig{file=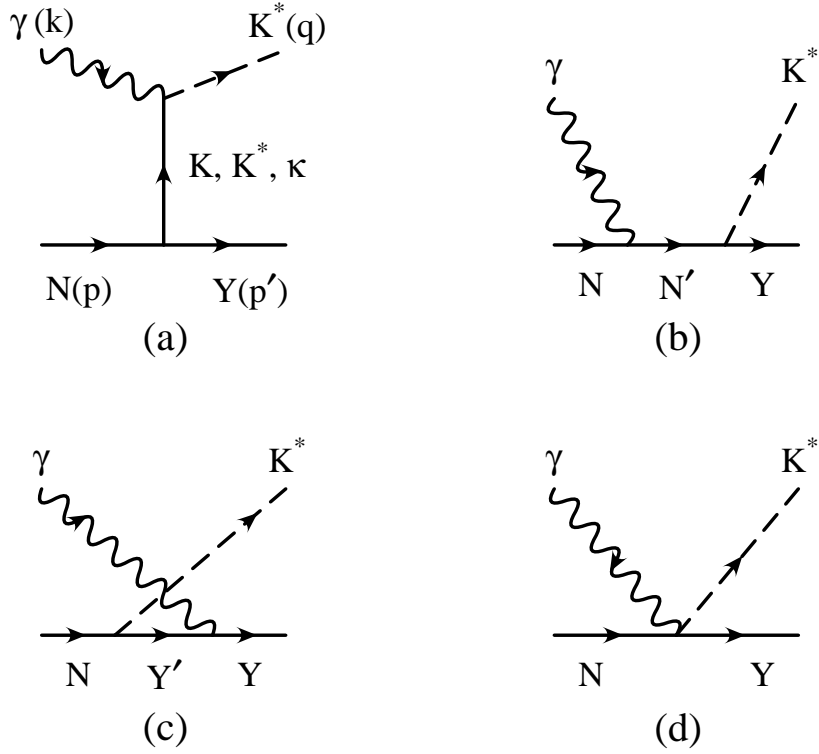, width=0.7\hsize}
\caption{Tree diagrams for $\gamma N \to K^* Y$ ($Y=\Lambda,\Sigma)$,
which include (a) $t$-channel exchanges, (b) $s$-channel $N$, $\Delta$,
(c) $u$-channel $\Lambda$, $\Sigma$, $\Sigma^*$, and (d) contact diagrams.}
\label{fig:diag}
\end{figure}

For $K^*$ photoproduction, we consider $t$-channel $K^*$, $K$,
$\kappa$ exchanges, $s$-channel $N$, $\Delta$,
and $u$-channel $\Lambda$, $\Sigma$, $\Sigma^*(1385)$ diagrams as shown in
Fig.~\ref{fig:diag}.
(The $t$-channel $K^*$ exchange and the contact diagram
Fig.~\ref{fig:diag}(d) are absent in $K^{*0}$ photoproduction.)
For the $t$-channel diagrams, which are expected to be dominant at small
$|t|$ region, the electromagnetic interactions are
\begin{eqnarray}
&& \mathcal{L}_{K^* K^*\gamma} = -ie A^\mu \left( K^{*-\nu}
K^{*+}_{\mu\nu} - K^{*-}_{\mu\nu} K^{*+\nu} \right),
\nonumber \\
&& \mathcal{L}_{K^*K\gamma} = g_{K^*K\gamma}^{}
\varepsilon^{\mu\nu\alpha\beta} \partial_\mu A_\nu \partial_\alpha
K_\beta^{*} \bar{K} + \mbox{ H.c.},
\nonumber \\
&& \mathcal{L}_{\kappa K^*\gamma} = e g_{\kappa K^*\gamma}^{} A^{\mu\nu}
\bar{\kappa} K^*_{\mu\nu} + \mbox{ H.c.},
\end{eqnarray}
where $A_\mu$ is the photon field, $A_{\mu\nu} = \partial_\mu A_\nu -
\partial_\nu A_\mu$, and $K^{*}_{\mu\nu} = \partial_\mu K^{*}_\nu -
\partial_\nu K^{*}_\mu$.
The decay width for $K^{*0} \to K^0\gamma$ ($K^{*\pm} \to  K^\pm\gamma$)
gives $g_{K^*K\gamma}^0 = -0.388$ GeV$^{-1}$ ($g_{K^*K\gamma}^c = 0.254$
GeV$^{-1}$).
The $\kappa$ meson couplings will be discussed later.

The $t$-channel hadronic interactions read
\begin{eqnarray}
\mathcal{L}_{K^* NY} &=& - g_{K^* NY}^{} \overline{N} \left( \gamma_\mu Y
 - \frac{\kappa_{K^* NY}^{}}{2M_N} \sigma_{\mu\nu} Y
\partial^\nu \right) K^{*\mu}
\nonumber \\ && \mbox{}
+ \mbox{ H.c.},
\nonumber \\
\mathcal{L}_{KNY} &=& -i g_{KNY}^{} \overline{N} \gamma_5 YK + \mbox{ H.c.},
\nonumber \\
\mathcal{L}_{\kappa NY} &=&
- g_{\kappa N Y}^{} \overline{N} Y \kappa
+ \mbox{ H.c.},
\label{lag2}
\end{eqnarray}
where $Y = \Lambda$ or $\bm{\tau} \cdot \bm{\Sigma}$.
The pseudoscalar coupling used for $\mathcal{L}_{KNY}$
is equivalent to the pseudovector coupling as the baryons are on-shell
in our case.
Then SU(3) relations are used to obtain
$g_{KN\Lambda}^{} = -13.24$ and $g_{KN\Sigma}^{} = 3.58$,
with $\alpha = 0.365$ and $g_{\pi NN}^2/4\pi = 14$.
For the $K^*$ couplings, the Nijmegen potential~\cite{RSY99-SR99} gives
$(g_{K^* NY}^{} = -4.26$, $\kappa_{K^* N Y}^{} = 2.66)$ for $Y=\Lambda$ and
$(-2.46,-0.47)$ for $Y=\Sigma$.
The Lagrangians and their coupling constants for the $s$- and
$u$-channel $N$, $\Delta$, $\Lambda$, $\Sigma$, and $\Sigma^*$ diagrams,
Figs.~\ref{fig:diag}(b,c), are fully discussed in
Refs.~\cite{OK06a,ONL04} and will not be repeated here. 
The contact diagrams, Fig.~\ref{fig:diag}(d), 
are required to have charge conservation in charged
$K^*$ production and can be calculated from the $K^*$ interaction
Lagrangian by minimal substitution.

One may also consider the axial-vector $K_1(1270)$ and
$K_1(1400)$ exchanges.
However, there are several comments for the interactions of the
axial-vector mesons.
Firstly, the $AV\gamma$ interaction like the $K_1\to K^*\gamma$ decay is an
anomalous interaction~\cite{Rosen63,KMOV00}, which does not exist in the
Bardeen subtracted anomalous action~\cite{KRS84}.
(See, however, Ref.~\cite{KM90b} for the hidden gauge approach.)
Although the $f_1 \to \rho\gamma/\phi\gamma$ decays are
seen, the other decays like $a_1 \to \rho\gamma/\omega\gamma$
have not been observed so far \cite{PDG04}.
Thus it is not yet clear whether the observed $f_1$ decays indicate the
existence of the $AV\gamma$ anomaly for the axial-vector meson nonet
or just reflect some peculiar internal structure of the $f_1$.
Secondly, the $K_1 NY$ couplings suffer from the lack of information.
(For the $a_1 NN$ coupling, see, e.g., Ref.~\cite{DBS84}.)
In addition, the large mass of $K_1$ mesons leads to an expectation
that the $K_1$ exchange contribution would be small.
Indeed, the total cross section data for $K^*\Lambda$ production
indicate suppressed contribution from high-spin meson exchanges in the
considered energy region~\cite{OK06a}.
Since there is no observation for the $K_1 \to K^*\gamma$ decay
so far, we leave the $K_1$ exchange for a future study.

Form factors are included to dress the vertices of the diagrams.
The following two forms are considered:
\begin{equation}
F_M(p_{\rm ex}^2) =
\frac{\Lambda^2 - M^2_{\rm ex}}{\Lambda^2 - p_{\rm ex}^2},
\quad
F_G(p_{\rm ex}^2) =
\frac{\Lambda^4}{\Lambda^4 + (p_{\rm ex}^2-M_{\rm ex}^2)^2},
\label{FF}
\end{equation}
where $M_{\rm ex}$ and $p_{\rm ex}$ are the mass and momentum
of the exchanged particle, respectively, and $\Lambda$ is the
cutoff parameter.
Including form factors can violate the charge conservation condition.
In fact, in $\gamma p \to K^{*+} \Lambda$, the sum of the $t$-channel
$K^*$ exchange, $s$-channel nucleon, and the contact term respects the
charge conservation when there is no form factor,
but they separately violate the condition \cite{OK06a}.
So introducing form factors depending on the exchanged particle can
easily break the charge conservation.
Following Ref.~\cite{DW01a}, charge conservation is restored by taking
the common form factor, $F = 1 - (1 - F_{K^*})(1-F_N)$,
for the three terms, where $F_{K^*}$ denotes the
$K^*$ exchange form factor, etc.
In $\gamma p \to K^{*0} \Sigma^+$, we have the same situation with the
$s$-channel nucleon and the $u$-channel $\Sigma$ terms,
and we take their common form factor as
$F = 1 - (1 - F_{N})(1-F_\Sigma)$.

\begin{figure}[t]
\centering
\epsfig{file=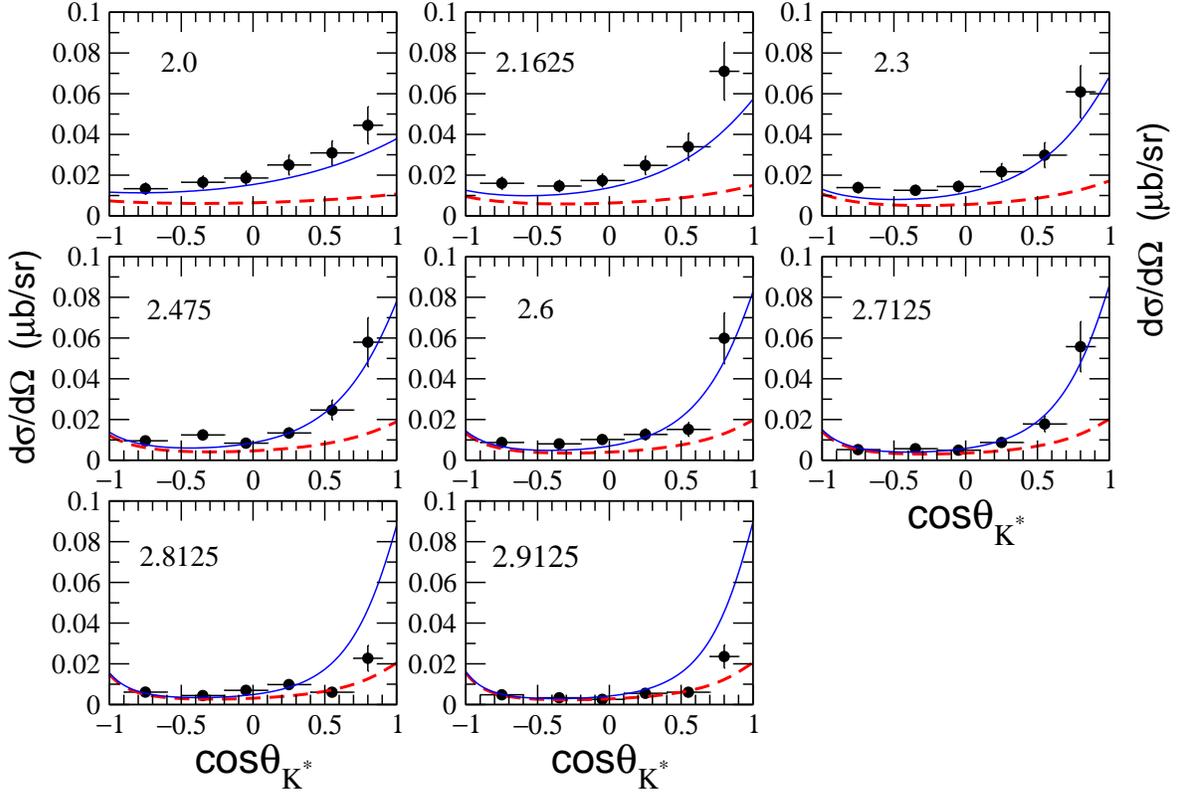, width=0.8\hsize, angle=-90}
\caption{Differential cross sections for $\gamma p \to K^{*0} \Sigma^+$.
The dashed and solid lines are the results for models (I) and
(II), respectively.
The number in each box denotes the photon energy $E_\gamma$ in GeV. The
preliminary data are from Ref.~\cite{HH05b}.}
\label{fig:sig-dif}
\end{figure}

In Ref.~\cite{OK06a}, considering all the diagrams of Fig.~\ref{fig:diag},
it was shown that the cross sections for $\gamma p
\to K^{*+}\Lambda$ could be well explained by the dominance of $K$ meson
exchange.
Here, the $t$-channel amplitudes have the form factors of the monopole
type $F_M$ with
$\Lambda_{K^*} = 0.9$ GeV and $\Lambda_K = \Lambda_\kappa = 1.1$ GeV.
The $s$- and $u$-channel form factors take the form of $F_G$ with
$\Lambda = 0.9$ GeV following Ref.~\cite{ONL04}.
In Ref.~\cite{OK06a}, $M_\kappa = 900$ MeV and $\Gamma_\kappa = 550$ MeV
were used following Ref.~\cite{BHS02}.
This is our model (I), where the $\kappa$ exchange was found to be small for
$K^*\Lambda$ production.
If we apply this model to $\gamma p \to K^{*0} \Sigma^+$, however,
we evidently underestimate the data as shown by the dashed lines
in Fig.~\ref{fig:sig-dif}.
This is consistent with the expectation with the $K$ exchange
dominance and indicates that the main production mechanisms of
$K^*\Lambda$ and $K^*\Sigma$ productions should be quite different.

In this paper, by observing the similarities in the differential cross
section data for $K^*\Lambda$ and for $K^*\Sigma$ productions, we propose
a different model where the scalar meson exchange plays a more important
role, especially in $K^*\Sigma$ case.
In fact, the mass and coupling constants of the $\kappa$ are not
firmly established, and model (I) uses
\begin{eqnarray}
|g_{\kappa K^*\gamma}^{c} g_{\kappa N \Lambda}| &=& 1.1 \mbox{ GeV}^{-1},
\nonumber \\
|g_{\kappa K^*\gamma}^{c} g_{\kappa N \Sigma}| &=& 0.7 \mbox{ GeV}^{-1},
\label{g_kap}
\end{eqnarray}
which are in the range of Refs.~\cite{BHS02,RSY99-SR99},
i.e., $|g_{\kappa K^*\gamma}^{c} g_{\kappa N \Lambda}^{}| = (1.0 \sim 1.2)$
and $|g_{\kappa K^*\gamma}^{c} g_{\kappa N \Sigma}^{}| = (0.6 \sim 0.8)$ in
GeV$^{-1}$ unit~\cite{OK06a}.
Also the SU(3) relation,
$g_{\kappa K^*\gamma}^{0} = -2 g_{\kappa K^*\gamma}^{c}$,
was used.
Because of the uncertainties in the couplings as well as in the
mass of the $\kappa$, we vary them within the acceptable ranges and
look for their values that reproduce the data for $K^*\Sigma$ photoproduction.

A successful description of the preliminary data of Ref.~\cite{HH05b} was 
achieved with $M_\kappa = 750$ MeV and the coupling constants
(\ref{g_kap}) by employing the form factor for the $\kappa$ exchange
in the form of $F_G$ with $\Lambda_\kappa = 1.2$ GeV,
while keeping the other production amplitudes as in model (I).
This is our model (II). We use $\Gamma_\kappa = 550$ MeV, whose
uncertainty, however, does not have significant influence.
The obtained results are given by the solid lines in Fig.~\ref{fig:sig-dif},
which imply that the off-shell $\kappa$ meson favors the form
factor in the form of $F_G$ over the mono-pole type $F_M$.
The main difference between the two form factors is that $F_G$ is
harder for small $|p_\kappa^2|$ and softer for large $|p_\kappa^2|$ compared
with $F_M$.
Therefore, microscopic studies on the behavior of the off-shell $\kappa$
meson couplings are highly desirable for understanding the internal
structure of the scalar mesons and the $\kappa$ meson exchange for $K^*$
photoproduction.
Since the scalar $\kappa$ meson exchange does not interfere with
the $K$ meson exchange, the unknown phases of the $\kappa$ meson couplings
(\ref{g_kap}) do not change our results. 
However, it should also be mentioned that there can be other choices for the
$\kappa$ meson parameters to describe $K^*\Sigma$ photoproduction.
For example, in model (II), by taking $M_\kappa = 900$ MeV with 
$|g_{\kappa K^*\gamma}^{c} g_{\kappa N \Sigma}^{}| = 1.2 \mbox{ GeV}^{-1}$
or $M_\kappa = 600$ MeV with
$|g_{\kappa K^*\gamma}^{c} g_{\kappa N \Sigma}^{}| = 0.4 \mbox{ GeV}^{-1}$,
we could obtain the results that are very close to the solid lines of
Fig.~\ref{fig:sig-dif}.
This shows that the uncertainties of the $\kappa$ meson parameters
\cite{PDG04} cannot be reduced by the current analyses on $K^*$ production,
and hence we do not make a fine tuning of the $\kappa$ parameters here.
In addition, in order to check whether such a role can be ascribed to a
more massive scalar meson, $K_0 (1430)$, we simply increased the
$\kappa$ mass to $1430$ MeV and found that its contribution is suppressed
due to the large mass.
Therefore, the $K^*$ photoproduction data can be used to
constrain the $\kappa$ meson parameters and a light scalar $\kappa$ meson
with $M_\kappa < 900$ MeV is favored.

\begin{figure}[t]
\centering
\epsfig{file=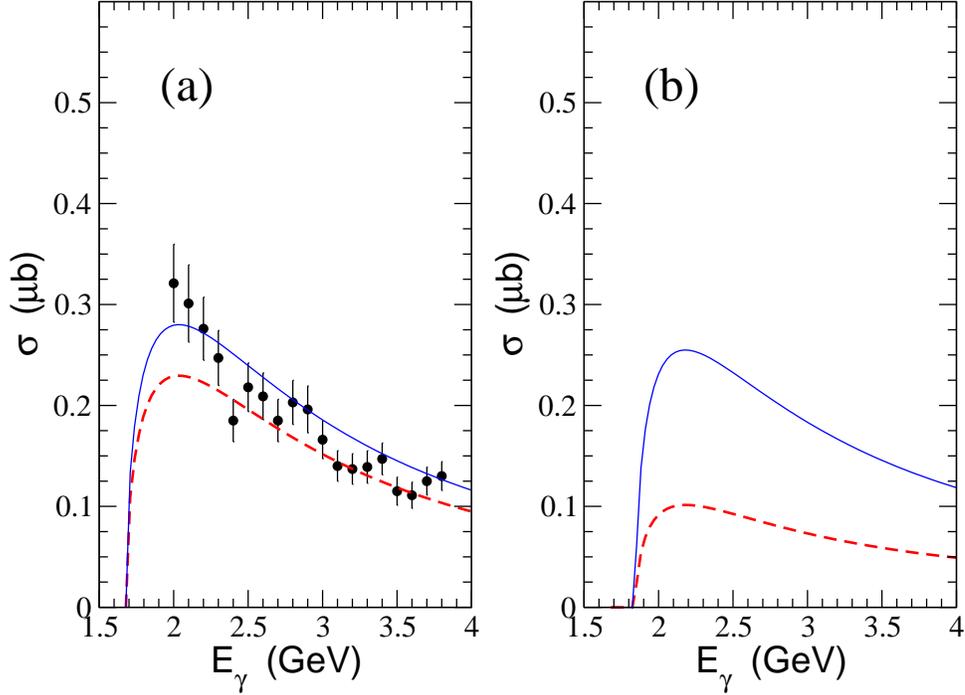, width=0.7\hsize, angle=-90}
\caption{Total cross sections for (a) $\gamma p \to K^{*+} \Lambda$ and
for (b) $\gamma p \to K^{*0} \Sigma^+$.
The dashed and solid lines are the results for models (I) and
(II), respectively. The data are from Ref.~\cite{GW06}.}
\label{fig:sig-tot}
\end{figure}

In model (II), we have shown that the scalar $\kappa$ meson exchange
might be crucial in the $K^*\Sigma$ production mechanisms.
Since the $\kappa$ meson parameters are different from
those of model (I), the previous results for $K^*\Lambda$
photoproduction should be re-examined.
We found that in model (II) the $K$ meson exchange is still
dominant for $K^*\Lambda$ production.
This is mainly due to the large value of $g_{KN\Lambda}^{}$.
Furthermore, since $\alpha \simeq 1.1$ for the scalar mesons
\cite{RSY99-SR99}, the coupling constant ratio of $\kappa$ exchange,
$R_\kappa \equiv (g^c_{\kappa K^*\gamma} g_{\kappa N\Lambda}^{} /
\sqrt2 g^0_{\kappa K^*\gamma} g_{\kappa N\Sigma}^{})^2 \simeq 0.3$,
implies a mild role of the $\kappa$ exchange in $K^{*}\Lambda$ production.
In SU(6) limit, $\alpha = 1$ for the scalar mesons and for the vector
couplings of the vector mesons, while $\alpha=2/5$ for the
other mesons \cite{SW65a}.
Thus, at least in this limit, the scalar meson is unique in  
giving a larger contribution to the $K^{*0}\Sigma^+$ channel than to
the $K^{*+}\Lambda$ channel, since $K^*$ exchange is absent in $K^{*0}$
production.
The obtained total cross sections for $K^*\Lambda$ production are given in
Fig.~\ref{fig:sig-tot} with those for $K^*\Sigma$ case.
This shows that the difference between model (I) and (II) for
$K^*\Sigma$ photoproduction is substantial, while it is small for
$K^*\Lambda$ case and is in the range of experimental errors.

\begin{figure}[t]
\centering
\epsfig{file=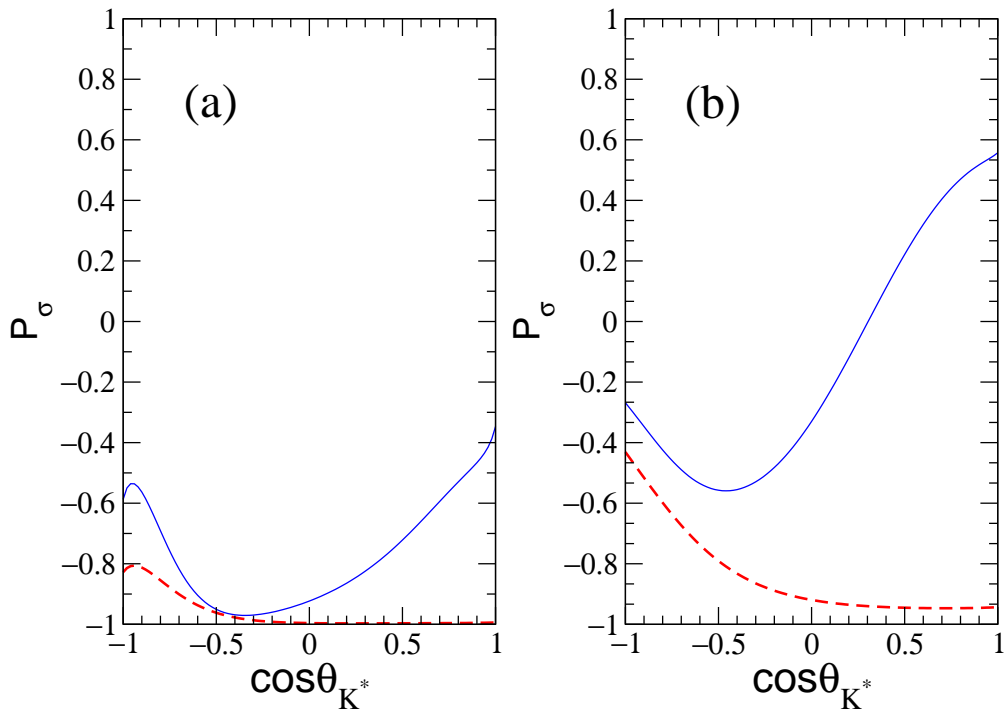, width=0.7\hsize, angle=-90}
\caption{Parity spin asymmetry $P_\sigma$ for
(a) $\gamma p \to K^{*+} \Lambda$ and for (b) $\gamma p \to K^{*0}
\Sigma^+$ at $E_\gamma = 3.0$ GeV.
Notations are the same as in Fig.~\ref{fig:sig-tot}.}
\label{fig:psigma}
\end{figure}

The scalar $\kappa$ meson has natural parity and the pseudoscalar
$K$ meson has unnatural parity.
The relative strength of the natural/unnatural $t$-channel exchanges
can be unambiguously estimated by measuring the parity
asymmetry~\cite{CSM68-SSW70},
\begin{equation}
P_\sigma \equiv \frac{d\sigma^N - d\sigma^U}{d\sigma^N + d\sigma^U} =
2 \rho_{1-1}^1 - \rho^1_{00},
\end{equation}
where $\rho$'s are the $K^*$ density matrix elements, and $d\sigma^N$
($d\sigma^U$) is the cross section from the natural (unnatural) parity
exchanges.
Therefore, we roughly expect that $P_\sigma$ is close to $-1$ when the kaon
exchange dominates, and its deviation from $-1$ shows the relative size of
the $\kappa$ and $K^*$ meson exchanges.
In order to avoid the contamination due to the $s$- and $u$-channel
amplitudes, it should be measured at relatively high energies and
at small scattering angles.
Shown in Fig.~\ref{fig:psigma} are the results for $P_\sigma$ at
$E_\gamma = 3.0$ GeV.
This shows the sensitivity of $P_\sigma$ on the scalar $\kappa$ meson
exchange, especially, in $K^{*0}\Sigma^+$ production since it excludes
natural-parity $K^*$ exchange.
Measuring the parity asymmetry is, therefore, highly required for
identifying the role of light $\kappa$ meson.
The same conclusion can be drawn for the photon beam asymmetry
$\Sigma_V\equiv (\rho^1_{11} + \rho^1_{1-1})/(\rho^0_{11} +
\rho^0_{1-1})$ \cite{CSM68-SSW70}.

In summary, we have investigated photoproduction mechanisms for
$K^* \Sigma$ and $K^* \Lambda$ within the tree level approximation,
especially focusing on the role driven by the scalar $\kappa$ meson exchange.
We found that the contribution from the light $\kappa$ meson with a mass
around $600 \sim 900$ MeV could be substantial for the $K^*\Sigma$
production, while it is supplementary in $K^*\Lambda$ production.
Therefore, $K^* \Sigma$ photoproduction provides a nice tool for studying
the controversial scalar $\kappa$ meson: specifically the parity asymmetry
and the photon beam asymmetry can be outstanding probes to separate the
$\kappa$ meson exchange in $K^*$ photoproduction, which
can be verified at current experimental facilities.


We are grateful to L. Guo, I.~Hleiqawi, and D.P. Weygand
for discussions and for providing us with their preliminary
data for $K^*$ photoproduction.
We thank T.-S.H. Lee and K. Nakayama for fruitful discussions.
Y.O. was supported by COSY Grant No. 41445282 (COSY-058).

\end{document}